\documentclass[twocolumn]{aastex62}

\usepackage{mathrsfs}
\usepackage{graphicx}
\usepackage{amsmath}

\received{\today}
\received{December 21, 2018}
\accepted{March 6, 2019}
\shorttitle{A Mean-Field Approach to Merging}
\shortauthors{Hozumi et al.}

\begin{document}

\title{A Mean-Field Approach to Simulating the Merging of Collisionless
Stellar Systems Using a Particle-Based Method}

\correspondingauthor{Shunsuke Hozumi}
\email{hozumi@edu.shiga-u.ac.jp}

\author{Shunsuke Hozumi}
\affil{Faculty of Education, Shiga University,
2-5-1 Hiratsu, Otsu, Shiga 520-0862, Japan}

\author{Masaki Iwasawa}

\author{Keigo Nitadori}
\affil{RIKEN Center for Computational Science,
7-1-26 Minatojima-minami-machi, Chuo-ku, Kobe, Hyogo 650-0047, Japan}

\begin{abstract}
We present a mean-field approach to simulating merging processes
of two spherical collisionless stellar systems.  This approach
is realized with a self-consistent field (SCF) method in which
the full spatial dependence of the density and potential of
a system is expanded in a set of basis functions for solving
Poisson's equation.  In order to apply this SCF method to a
merging situation where two systems are moving in space, we
assign the expansion center to the center of mass of each
system, the position of which is followed by a mass-less
particle placed at that position initially.  Merging simulations
over a wide range of impact parameters are performed using
both an SCF code developed here and a tree code.  The results
of each simulation produced by the two codes show excellent
agreement in the evolving morphology of the merging systems
and in the density and velocity dispersion profiles of the
merged systems.  However, comparing the results generated
by the tree code to those obtained with the softening-free
SCF code, we have found that in large impact parameter cases,
a softening length of the Plummer type introduced in the tree
code has an effect of advancing the orbital phase of the two
systems in the merging process at late times.  We demonstrate
that the faster orbital phase originates from the larger
convergence length to the pure Newtonian force.  Other
application problems suitable to the current SCF code
are also discussed.
\end{abstract}

\keywords{galaxies: evolution --- galaxies: kinematics and dynamics
--- methods: numerical}

\section{Introduction}\label{sec:introduction}
$N$-body simulation is an indispensable method for studying
astronomical objects whose constituents interact gravitationally
with one another.  In order to reveal detailed structures of
a target object which is modeled with $N$ particles, we need
to invest in it as many particles as possible.  However, extremely
large $N$-body simulations are prohibitively time-consuming
because the number of force calculation per time step increases
explosively as in $\mathcal{O}(N^2)$ with increasing $N$, if
the force exerted on a given particle is calculated by summing
up the mutual gravitational forces over all other particles.
This disadvantage has been ameliorated by adopting, e.g., a
tree algorithm \citep{bh86} which reduces the number of force
calculation to $\mathcal{O}(N\log N)$.  Regarding this algorithm,
a recently developed numerical library called FDPS has facilitated
the implementation of a tree code running with a sufficiently
high speed on a massively distributed-memory parallel computer
\citep{iwasawa16,namekata18}, which will promote simulations with
an ever-larger number of particles.

When an $N$-body method is applied to collisionless systems such
as galaxies and clusters of galaxies, we should pay attention to
the difference between a system composed of $N$ particles and its
corresponding genuine collisionless system.  In a strict sense,
an $N$-body system in which the forces among particles are
calculated, in essence, using pair-wise interactions is not
collisionless.  This is because the term ``collisionless" means
that particles interact with the mean force field generated
by themselves, and so, the forces do not depend on the relative
coordinates of pair-wise particles.  However, by terminating an
$N$-body simulation within a period much shorter than the two-body
relaxation time even in that way of pair-wise force calculation,
we are allowed to regard the system as collisionless in a
practical sense.  On the other hand, a mean-field approach
ensures the collisionless nature at least formally.  In reality,
if a mean field is represented by a particle-based method,
Poisson fluctuations induced by the finite number of particles
cause collisional effects equivalent to two-body relaxation
\citep{larsjosh90}.  Nevertheless, such an approach is still
desirable for collisionless stellar systems from some points
of view.  For example, once the force field is given, the
orbits of all particles can be calculated independently,
and thereby one $N$-body problem is reduced to $N$ one-body
problems, as has been pointed out by \citet{ho92}.  This
approach is thus well suited to parallel computation as
shown by \mbox{\citet{hsb95}}, so that extremely large-$N$
simulations will be made feasible.  In addition, if the mean
field is known at each time step, we can trace the orbits of
particles starting from arbitrary positions in phase space,
which enables us to reproduce phase space itself at a given
time, as has been demonstrated by \citet{sh97}.

We can realize a mean-field approach to simulating collisionless
stellar systems using expansion techniques for solving Poisson's
equation.  Above all, a self-consistent field (SCF) method
termed by \citet{ho92}, the idea of which dates back to that
of \citet{cb72,cb73}, is considerably useful.  The essence of
the SCF method consists in expanding the density and potential
of a system in a set of basis functions.  In this method, the
expansion of the full spatial dependence makes the cpu cost
proportional to particle number $N$.  In addition, ideal
load-balancing is easily achieved on a massively parallel
machine owing to the perfect scalability inherent in the
collisionless nature.

One of the disadvantages intrinsic to SCF methods is the
limitation in the range of applicability because the expansion
center is needed at every time step.  By virtue of this inconvenience,
SCF methods have been applied only to stellar dynamical problems
in which systems of interest are isolated and fixed in space with
the center of mass being immovable such as violent relaxation of
stellar systems \citep{hh95,hfk96}, halo evolution \citep{wk02},
and disk evolution \citep{hh05,sh12}.  However, moving systems are
ubiquitous in the real Universe, as gravitationally interacting
events such as tidal encounter, collision and merging are commonly
observed.  In particular, merging is the fundamental process
of hierarchical structure formation in the universe dominated
by cold dark matter.  Even after galaxies have fully grown up
through hierarchical merging, they encounter or collide, and
frequently result in merging to each other.  Accordingly, many
simulations have been devoted to reproducing the characteristic
appearance of those tails, bridges, and antennae which are
indicative of galaxy interactions \citep[e.g.,][]{tt72,lt76,ts77,fs82,
ws92,howard93,josh98,nb03}.  Therefore, if SCF methods
can be applied to merging systems, we will be able to understand
in detail the features produced by merging processes and the
properties of merged systems using the huge number of particles.
As an instance, if the cusp found in the central region of a dark
matter halo \citep{nfw96,nfw97} arises from the merging of clumps
as suggested by \citet{fm97}, an SCF technique might help elucidate
the origin of the power-law nature of the cusp that continues
almost down to the center by assigning a tremendously large
number of particles to each clump.

In conventional $N$-body methods, a softening length is
introduced explicitly in order to avoid numerical divergences
of the mutual force when two particles pass close to each other.
On the other hand, an SCF method includes implicit softening for
the force calculation in the sense that the force resolution is
limited to a certain degree of precision because the expansion
terms of the density and potential of a system are inevitably
truncated at the finite numbers in a set of basis functions.
However, the non-existence of an explicit scale length in
interaction forces suggests that an SCF method may be able
to describe the evolution of a stellar system under the pure
\mbox{Newtonian} force law more faithfully than an explicitly
softening-dependent method.  In general, gravitational
softening affects more severely the dynamics of so-called cold,
rotation-dominated systems like galaxy disks \mbox{\citep{miller71,
miller74, es95}} than that of hot systems like elliptical galaxies
which are supported by velocity dispersion.  In a sense, the
orbital motion of two merging spherical galaxies is dynamically
cold, even though each galaxy is a hot elliptical-like system.
Consequently, the softened gravity might influence a merging
process.  In fact, we will uncover the effects of a softening
length on the orbital phase of two merging systems.

In this paper, we present merging simulations of two spherical
collisionless stellar systems on the basis of a mean-field
approach which is realized using an SCF code developed here.
In Section \ref{sec:models}, models and initial settings employed
are described.  In Section~\ref{sec:methods}, we explain how
an SCF method is applied to merging simulations, along with
the details of tree code simulations which are used not only
for comparison but for making clear the effects of a softening
length.  Results are shown in Section~\ref{sec:results}.
We discuss the effects of a softening length, computation time,
and application problems of the SCF code for merging simulations.
Conclusions are given in Section~\ref{sec:discussion}.

\section{Models}\label{sec:models}
\begin{table}[b!]
\centering
\caption{Initial Cartesian coordinates of the centers of the two systems.}
\label{tab:cases}
\begin{tabular}{crr}
\tablewidth{0pt}
\hline
\hline
Name & System 1 & System 2\\
\hline
Case 0 & $(-5,\,0,\,0)$ & $(5,\,0,\,0)$\\
Case 1 & $(-5,\,-1,\,0)$ & $(5,\,1,\,0)$\\
Case 2 & $(-5,\,-2,\,0)$ & $(5,\,2,\,0)$\\
Case 3 & $(-5,\,-3,\,0)$ & $(5,\,3,\,0)$\\
Case 4 & $(-5,\,-4,\,0)$ & $(5,\,4,\,0)$\\
Case 5 & $(-5,\,-5,\,0)$ & $(5,\,5,\,0)$\\
\hline
\end{tabular}
\end{table}

We use two identical spherical King models \citep{king66} for merging
simulations.  The distribution function (DF) of a King model
is expressed by
\begin{equation}
f(\mathcal{E})=\left\{
\begin{array}{ll}
\displaystyle\frac{\rho_1}{{(2\pi\sigma^2)}^{3/2}}(e^{\mathcal{E}/\sigma^2}-1) &
             (\mathcal{E} > 0),\\
0 & (\mathcal{E} \le 0),
\end{array}\right.
\label{eq:king_DF}
\end{equation}
where $\rho_1$ and $\sigma$ are some constants which have
the dimensions of density and velocity dispersion, respectively.
In Equation~(\ref{eq:king_DF}),
$\mathcal{E}$ is the relative energy defined by
\begin{equation}
\mathcal{E} = \Psi(r) -\frac{1}{2}({v_x}^2+{v_y}^2+{v_z}^2),
\label{eq:rel_energy}
\end{equation}
where $\Psi$ is the relative potential, $r$ is the radial distance
from the center, and $v_x, v_y$, and $v_z$ are the three \mbox{Cartesian}
components of the velocity vector.  The potential of the system,
$\Phi$, is related to $\Psi$ as
\begin{equation}
\Phi(r)=-\Psi(r)-\frac{GM}{r_t},
\label{eq:potential}
\end{equation}
where $G$ is the gravitational constant, $M$ is the total
mass, and $r_t$ is the tidal radius that makes the system
finite with $\Psi(r_t)=0$.

We use the dimensionless central potential $W_0=3$, where $W_0$
is defined by $W_0=\Psi(0)/\sigma^2$, eliminating the arbitrariness
of the DF shown in Equation~(\ref{eq:king_DF}).  In this case, we
have $r_t=4.70\,r_0$ with $r_0$ being the core radius such that
\begin{equation}
r_0=\sqrt{\frac{9\sigma^2}{4\pi G\rho_0}},
\label{eq:r_core}
\end{equation}
where $\rho_0$ is the central density.

The King model is realized with $N=10,000,584$ particles of equal
mass.  A dimensionless system of units is chosen such that $G=1$,
$M=1$, and $r_0=1$.

We set the $xy$-plane to be the orbital plane of the two systems.
At the beginning, each system is put on either side of the
$y$-axis.  The system placed in the region with $x < 0$ and
that in the region with $x>0$ are called System~1 and System~2,
respectively.  We study 6 cases listed in Table \ref{tab:cases}.
The initial velocity components, $(v_{0,x},\, v_{0,y},\, v_{0,z})$,
of the centers of mass of System~1 and System~2 are, respectively,
$(0.2,\, 0,\, 0)$ and $(-0.2,\, 0,\, 0)$.

\section{Methods}\label{sec:methods}

\subsection{SCF Method}
Suppose that there exists a basis set consisting of density and
potential basis functions denoted by $\rho_{nlm}(\textbf{\textit{r}})$
and $\Phi_{nlm}(\textbf{\textit{r}})$, respectively, where $n$
is a discrete number in the radial direction, $l$ and $m$
are corresponding quantities in the angular directions, and
$\textbf{\textit{r}}$ is the position vector.
Each pair of the basis functions satisfies Poisson's equation
written by
\begin{equation}
\nabla^2\Phi_{nlm}(\textbf{\textit{r}})=4\pi
G\rho_{nlm}(\textbf{\textit{r}}),
\end{equation}
and the density and potential basis functions have the
bi-orthonormality represented by
\begin{equation}
\int \rho_{nlm}(\textbf{\textit{r}})\Phi_{n'l'm'}(\textbf{\textit{r}})
\,d\textbf{\textit{r}}=\delta_{nn'}\delta_{ll'}\delta_{mm'},
\label{eq:orthonormal}
\end{equation}
where $\delta_{kk'}$ is the Kronecker delta defined by
$\delta_{kk'}=0$ for $k\ne k'$ and $\delta_{kk'}=1$ for $k=k'$.
Then, Poisson's equation for an isolated self-gravitating system
is solved by expanding its density and potential in that basis set
with respect to its center of mass.  As a consequence, we find
\begin{equation}
\rho(\textbf{\textit{r}})=\sum_{n,l,m}A_{nlm}(t)
\rho_{nlm}(\textbf{\textit{r}})
\label{eq:rho_basis}
\end{equation}
and
\begin{equation}
\Phi(\textbf{\textit{r}})=\sum_{n,l,m}A_{nlm}(t)
\Phi_{nlm}(\textbf{\textit{r}}),
\label{eq:pot_basis}
\end{equation}
where $A_{nlm}(t)$ are the expansion coefficients at time $t$.

Once a basis set is given, the potential basis functions,
$\Phi_{nlm}(\textbf{\textit{r}})$, are operated to the
underlying density distribution, $\rho(\textbf{\textit{r}})$,
which can be expressed by Equation~(\ref{eq:rho_basis}), and then,
$A_{nlm}(t)$ are obtained with the help of the bi-orthonormality
between $\rho_{nlm}(\textbf{\textit{r}})$ and
$\Phi_{nlm}(\textbf{\textit{r}})$ indicated by
Equation~(\ref{eq:orthonormal}) such that
\begin{eqnarray}
\int \rho(\textbf{\textit{r}})\Phi_{nlm}(\textbf{\textit{r}})
\,d\textbf{\textit{r}} & = & \int
\sum_{n',l',m'}A_{n'l'm'}(t)\,\rho_{n'l'm'}(\textbf{\textit{r}})
\nonumber\\
& & \times\Phi_{nlm}(\textbf{\textit{r}})\,d\textbf{\textit{r}} \nonumber\\
& = & \sum_{n',l',m'}A_{n'l'm'}(t)\delta_{n'n}\delta_{l'l}\delta_{m'm}\nonumber\\
& = & A_{nlm}(t).
\label{eq:exp_coef}
\end{eqnarray}
On the other hand, since an $N$-body system is a collection of
discrete mass-points, we find
\begin{eqnarray}
A_{nlm}(t)& =& \int \rho(\textbf{\textit{r}})\,\Phi_{nlm}(\textbf{\textit{r}})
\, d\textbf{\textit{r}} \nonumber\\
& = & \int \sum_k m_k\, \delta(\textbf{\textit{r}}-\textbf{\textit{r}}_k)
\, \Phi_{nlm}(\textbf{\textit{r}}) \, d\textbf{\textit{r}} \nonumber\\ 
& = & \sum_k m_k\, \Phi_{nlm}(\textbf{\textit{r}}_k),
\label{eq:discrete_exp_coef}
\end{eqnarray}
where $m_k$ is the mass of the $k$-th particle in the system,
$\textbf{\textit{r}}_k$ is its position vector at time $t$, and
$\delta(\textbf{\textit{r}})$ is Dirac's delta function.

We can therefore compute $\Phi(\textbf{\textit{r}})$
from Equations~(\ref{eq:pot_basis}) and (\ref{eq:discrete_exp_coef}).
This potential then yields the acceleration,
$\textbf{\textit{a}}(\textbf{\textit{r}})$,
by differentiating Equation~(\ref{eq:pot_basis}) with respect to
$\textbf{\textit{r}}$, leading to
\begin{equation}
\textbf{\textit{a}}(\textbf{\textit{r}})
=-\sum_{n,l,m}A_{nlm}(t)\nabla\Phi_{nlm}(\textbf{\textit{r}}),
\label{eq:acc_basis}
\end{equation}
where $\nabla\Phi_{nlm}(\textbf{\textit{r}})$ can be calculated
analytically in advance.  The system is evolved forward in time
by a time step $\Delta t$ with a suitable integration scheme using
Equation~(\ref{eq:acc_basis}).  In this way, we can follow the
evolution of the system by iterating this procedure.

In contrast with isolated systems, two interacting systems
like those studied here are moving in space, which implies
that the center of mass (CM) of each system is not fixed but 
moving with time.  It thus follows that we need to locate
the expansion centers of the two systems at every time step.
For this purpose, when starting a simulation, we first put
a mass-less particle at the CM of each system as a guide of
the expansion center, and trace the position of that particle
by summing over all two-body interactions of it with the
self-gravitating particles.  That is, the acceleration of
the mass-less particle in System~$i$ ($i=1$ or 2),
$\textbf{\textit{a}}(\textbf{\textit{r}}_{{\rm CM},i})$,
is calculated as
\begin{equation}
\textbf{\textit{a}}(\textbf{\textit{r}}_{{\rm CM},i}) =-\sum_{k}
\frac{Gm_k(\textbf{\textit{r}}_k-\textbf{\textit{r}}_{{\rm CM},i})}
{{|\textbf{\textit{r}}_k-\textbf{\textit{r}}_{{\rm CM},i}|}^3},
\label{eq:accel_cm}
\end{equation}
where $\textbf{\textit{r}}_{{\rm CM},i}$ is the position vector
of the mass-less particle in System~$i$, and $\textbf{\textit{r}}_k$
stands for the position vector of the $k$-th self-gravitating particle
with the mass being $m_k$ in the total system.  Then, we calculate the
self-gravity of each system by expanding the density and potential
of the corresponding system around its expansion center.
The acceleration of a particle in System~$i$ due to the self-gravity,
$\textbf{\textit{a}}_{i\rightarrow i}(\textbf{\textit{r}})$,
is represented by
\begin{equation}
\textbf{\textit{a}}_{i\rightarrow i}(\textbf{\textit{r}}) =\sum_{n,l,m}
A_{nlm,i}(t)\nabla\Phi_{nlm}(\textbf{\textit{r}}
-\textbf{\textit{r}}_{{\rm CM},i}),
\end{equation}
where $A_{nlm,i}(t)$ are the expansion coefficients of
System~$i$ at time $t$ with the expansion center being
$\textbf{\textit{r}}_{{\rm CM},i}$.  Here, we assume that the
same basis set is used for Systems~1 and 2, since we deal with
the merging of two identical King models.  Next, the interaction
forces exerted on the particles in one system by those in the
other system are evaluated by applying the expansion coefficients
of the other system to the forces at their positions.  That is,
the acceleration of a particle in System~$i$ caused by
the force field of System~$j$ ($i\neq j$), 
$\textbf{\textit{a}}_{j\rightarrow i}(\textbf{\textit{r}})$,
is given by
\begin{equation}
\textbf{\textit{a}}_{j\rightarrow i}(\textbf{\textit{r}}) =\sum_{n,l,m}
A_{nlm,j}(t)\nabla\Phi_{nlm}(\textbf{\textit{r}}
-\textbf{\textit{r}}_{{\rm CM},j}),
\label{eq:accel_21}
\end{equation}
where $A_{nlm,j}(t)$ are the expansion coefficients of
System~$j$ at time $t$ with the expansion center being
$\textbf{\textit{r}}_{{\rm CM},j}$.

\begin{figure*}[ht]
\centerline{\includegraphics[width=16.5cm]{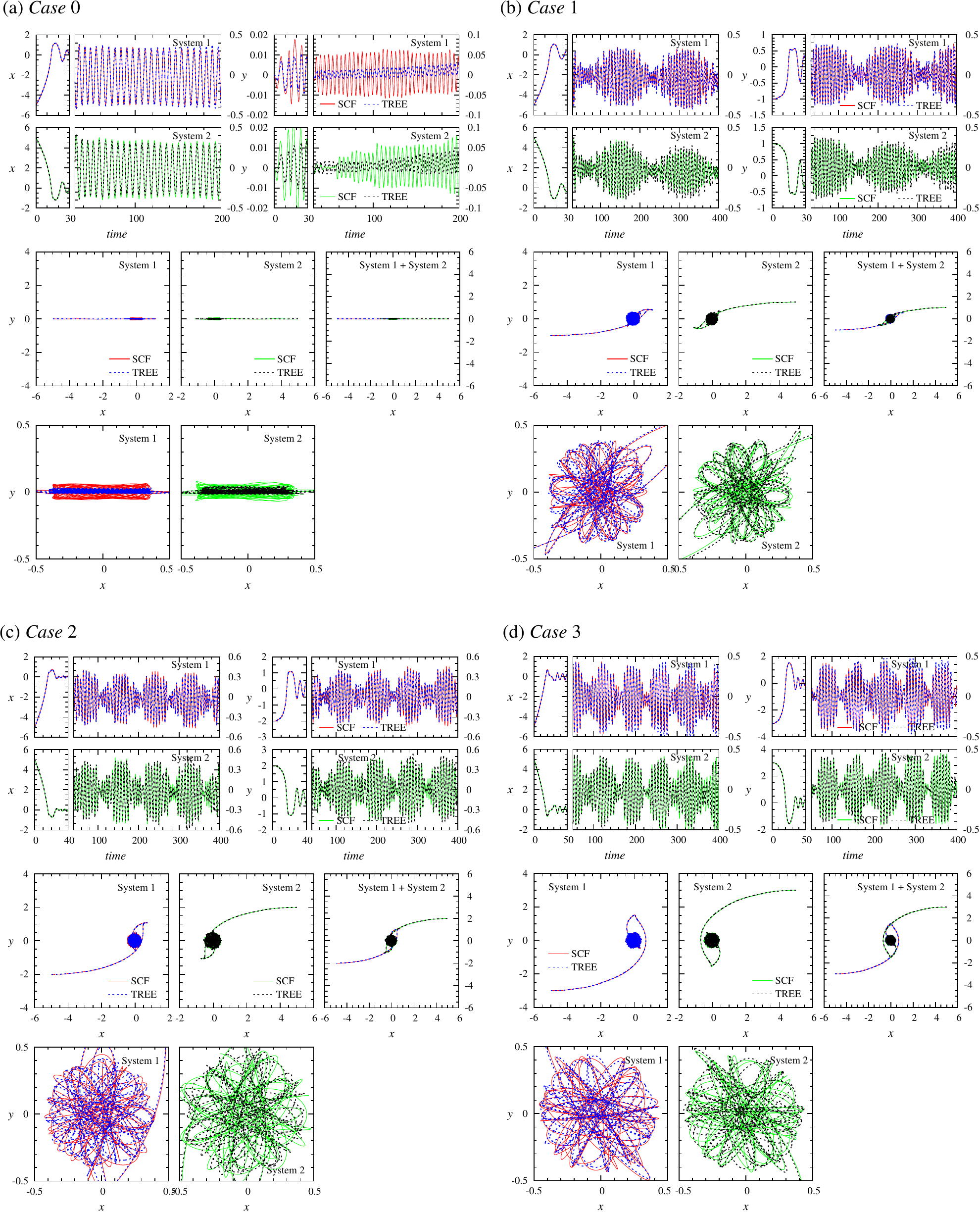}}
\caption{Motions of the centers of mass (CM) of Systems~1
and 2 for (a) Case~0 to (f) Case~5.  In each case, in the
first-row panels, the top panels show the $x$ (left two panels)
and $y$ (right two panels) coordinates of the CM of System~1
as a function of time, while the bottom panels present those
of System~2.  The second-row panels display the orbit of the
CM of System~1 (left panel), that of System~2 (middle panel),
and that of Systems~1 and 2 (right panel).  In the third-row
panels, the left panel is the enlargement version of the left
panel in the second row, while the right panel is that of the
middle panel in the second row.  The red and green solid lines
represent the CM of System~1 and that of System~2, respectively,
for the SCF simulation, while the blue and black dashed lines
denote the CM of System~1 and that of System~2, respectively,
for the tree code simulation.  In the SCF simulation, the
interaction forces are calculated using Equation~(\ref{eq:accel_21}).}
\label{fig:cases0-3}
\end{figure*}

For large impact parameters, however, we have found that the
calculation method based on Equation~(\ref{eq:accel_21}) bears
large errors.  Instead, we expand a system of interest with
respect to the CM of the total system, and apply the expansion
coefficients of that system to the forces suffered by the particles
in the other system.  Thus, instead of Equation~(\ref{eq:accel_21}),
we use
\begin{equation}
\textbf{\textit{a}}_{j\rightarrow i}(\textbf{\textit{r}}) =\sum_{n,l,m}
{A^{(\rm CM)}}_{nlm, j}(t)\nabla\Phi_{nlm}(\textbf{\textit{r}}
-{\textbf{\textit{r}}}_{\rm CM}),
\label{eq:accel_tot_cm}
\end{equation}
where ${A^{(\rm CM)}}_{nlm,j}(t)$ are the expansion coefficients
calculated by expanding System~$j$ around the CM of the
total system, ${\textbf{\textit{r}}}_{\rm CM}$.  In reality,
the conservation of linear momentum leads to
${\textbf{\textit{r}}}_{\rm CM}=0$, and ${A^{(\rm CM)}}_{nlm,j}(t)$
are identical to ${A^{(\rm CM)}}_{nlm,i}(t)$ ($i\neq j$), since
the two identical King models are set up initially in a configuration
symmetric about the origin.  In these cases, again, we use the same
basis set for the interaction forces between the two systems as that
used for the self-gravity.  After all, the acceleration of a particle
in System~$i$,
$\textbf{\textit{a}}_i(\textbf{\textit{r}})$, is calculated as
\begin{equation}
\textbf{\textit{a}}_i(\textbf{\textit{r}})=
\textbf{\textit{a}}_{i\rightarrow i}(\textbf{\textit{r}})+
\textbf{\textit{a}}_{j\rightarrow i}(\textbf{\textit{r}})\quad (i \ne j).
\end{equation}

\setcounter{figure}{0}
\begin{figure*}[th]
\centerline{\includegraphics[width=16.5cm]{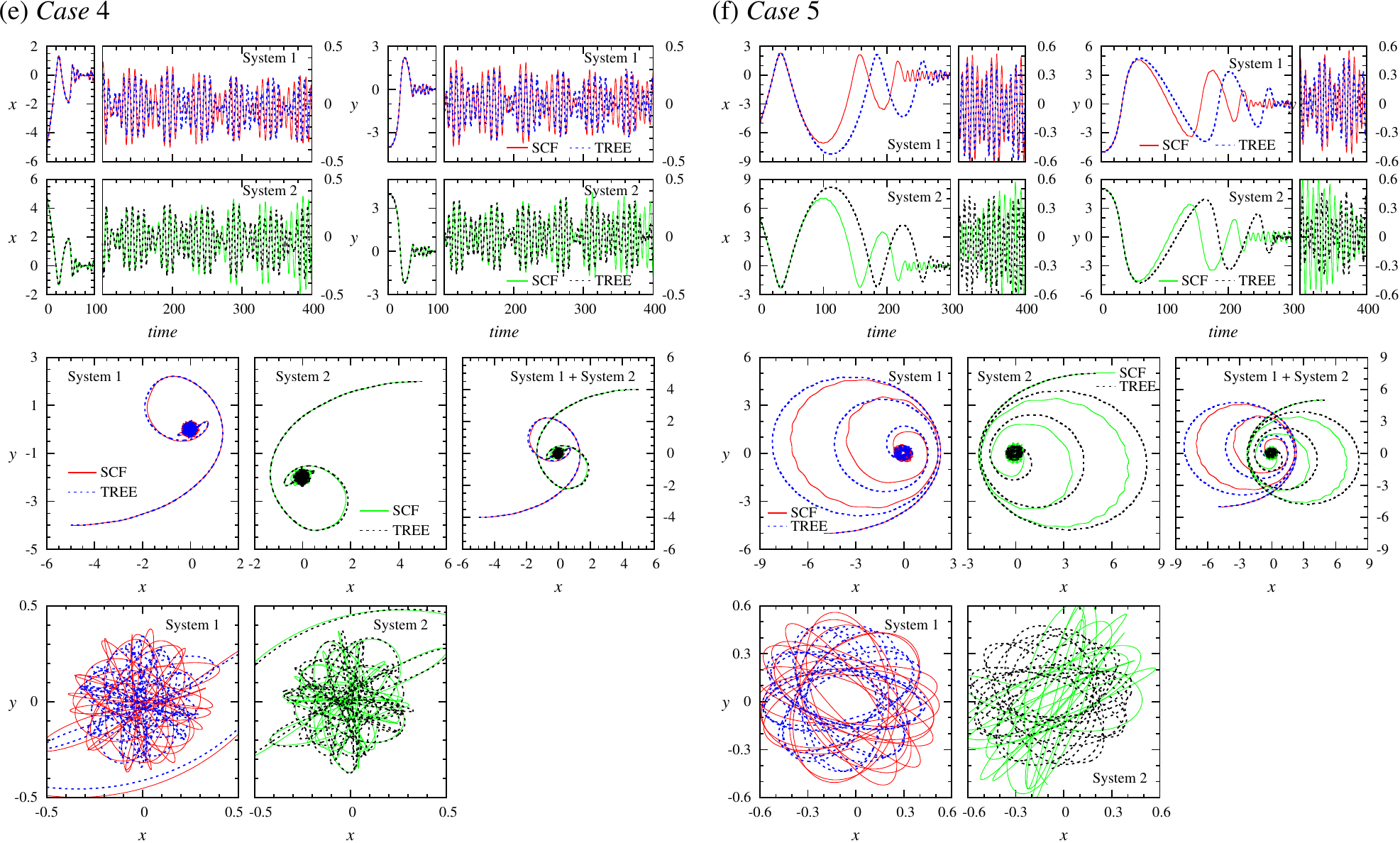}}
\caption{Continued.}
\label{fig:cases4-5}
\end{figure*}

\setcounter{figure}{1}
\begin{figure}[h]
\vspace{5ex}
\centerline{\includegraphics[width=8cm]{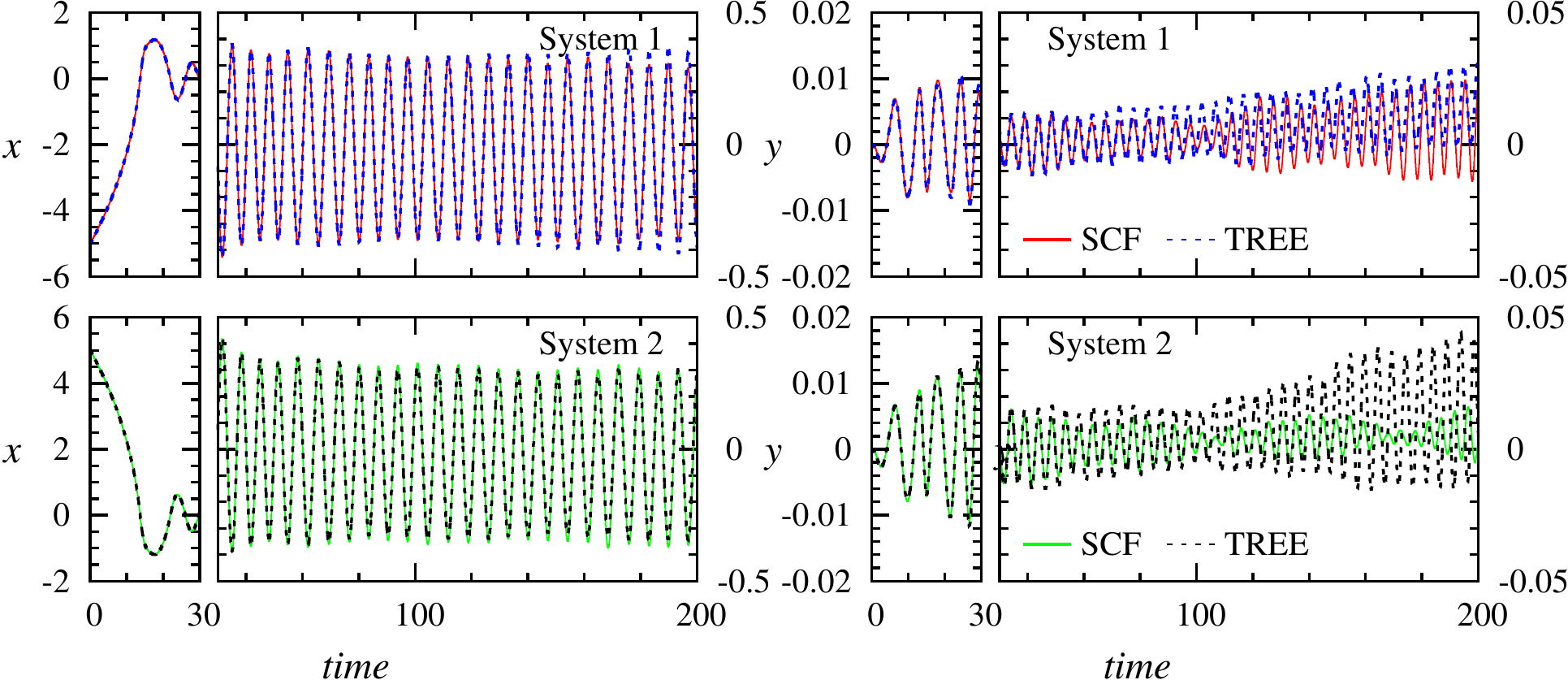}}
\caption{Same as in the first-row panels of Figure~\ref{fig:cases0-3}
(a) for the Case~0 simulations.  In the SCF simulation, the same
softening length as that adopted in the tree code one was employed
to calculate the orbits of the test particles placed at the centers
of mass of Systems~1 and 2.}
\label{fig:SCF_eps}
\end{figure}

Since the King model has a core structure, the basis functions
suitable for the present study are those which are not cuspy
but cored.  Accordingly, we select Clutton-Brock's basis set
\citep{cb73} which is constructed on the basis of the Plummer
model \mbox{\citep{plummer11}}.  The exact functional forms of the basis
set employed are given in Appendix \ref{app:A}.  The scale length
of the basis functions, $a$, is set to be $a=1.15$ so as to represent
the density distribution of the King model with the lowest order
density term, $\rho_{000}$, that is identical to the density of
the \mbox{Plummer} model, as closely as possible.  We do not change
the value of $a$ even in calculating the interaction forces between
the two systems.

We carried out convergence tests to determine the numbers of
the expansion terms.  After some trials, we adopt $n_{\rm max}=16$
and $l_{\rm max}=m_{\rm max}=10$, where $n_{\rm max}$ is the
maximum number of expansion coefficients in the radial direction,
and $l_{\rm max}$ and $m_{\rm max}$ are the maximum numbers
of those in the angular directions.  For the interaction forces
that are calculated using the CM of the total system as the
expansion center, we adopt $n_{\rm max}=28$ and
$l_{\rm max}=m_{\rm max}=28$ with the value of $a$ intact
for the same basis set, while keeping $n_{\rm max}=16$ and
$l_{\rm max}=m_{\rm max}=10$ for the calculations of the
self-gravitating forces.  Regarding the integration scheme
to solve the equations of motion for the self-gravitating
and mass-less particles, we employ a time-centered leapfrog
method \citep{press86} with a fixed time step of $\Delta t=0.01$.
As a result, the energy was conserved to better than 0.0882\%
for all the simulations.

\begin{figure}[h]
\vspace{5ex}
\centerline{\includegraphics[width=8cm]{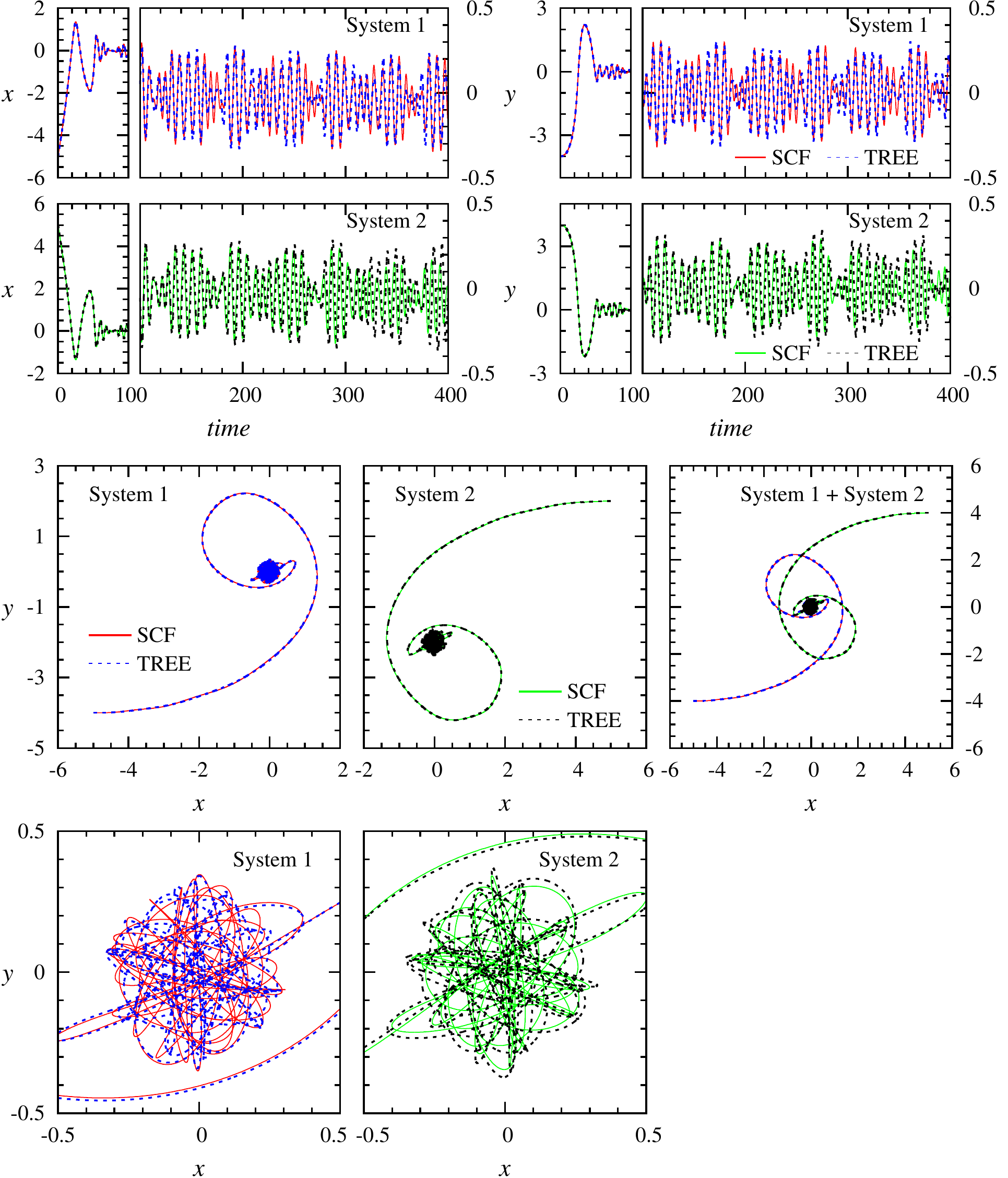}}
\caption{Same as in Figure~\ref{fig:cases4-5} (e) for the Case~4
simulations, but the interaction forces are calculated on
the basis of Equation~(\ref{eq:accel_tot_cm}) for the SCF simulation.}
\label{fig:case4_int}
\end{figure}

\subsection{Tree Method}
The same initial setups as those adopted in the SCF simulations
are used for simulations with a hierarchical tree algorithm
\citep{bh86} in order to evaluate the results produced
by the SCF code.  In so doing, we utilize an FDPS library
written in a C++ language \citep{iwasawa16} which has been
highly tuned for a massively distributed-memory parallel machine.
In the tree approach, forces are expanded up to quadrupole order,
and computed with an opening angle of 0.5, unless otherwise
mentioned.  We choose a gravitational softening of the Plummer
type to prohibit numerical divergences when two particles pass
close to each other, and set the softening length, $\varepsilon$,
to be the mean interparticle separation within the half-mass
radius, $\varepsilon=0.0074$, again unless otherwise mentioned.
In accordance with the SCF simulations, we put a mass-less
particle at the center of each system to compare its orbit
with that obtained using the SCF code.  For the mass-less
particles, the same softening length, $\varepsilon=0.0074$,
is adopted.  A time-centered leapfrog method is used with
a fixed time step of $\Delta t=0.01$.  As a result, all the
tree code simulations showed the relative energy error to be
smaller than 0.0174\%.

\begin{figure}[h]
\vspace{5ex}
\centerline{\includegraphics[width=8cm]{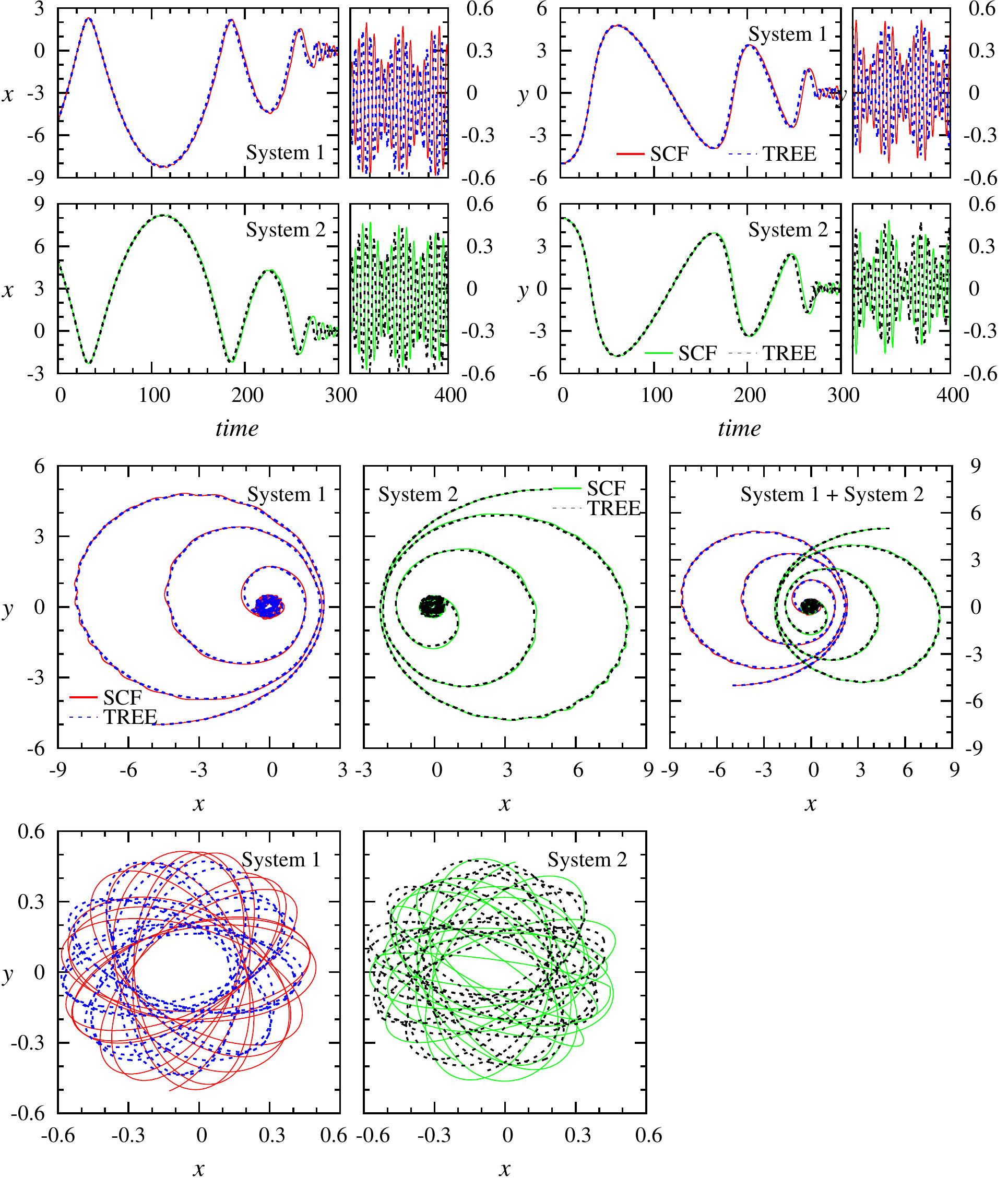}}
\caption{Same as in Figure~\ref{fig:cases4-5} (f) for the Case~5
simulations, but the interaction forces are calculated on
the basis of Equation~(\ref{eq:accel_tot_cm}) for the SCF simulation.}
\label{fig:case5_int}
\end{figure}

\begin{figure*}[h]
\centerline{\includegraphics[width=14cm]{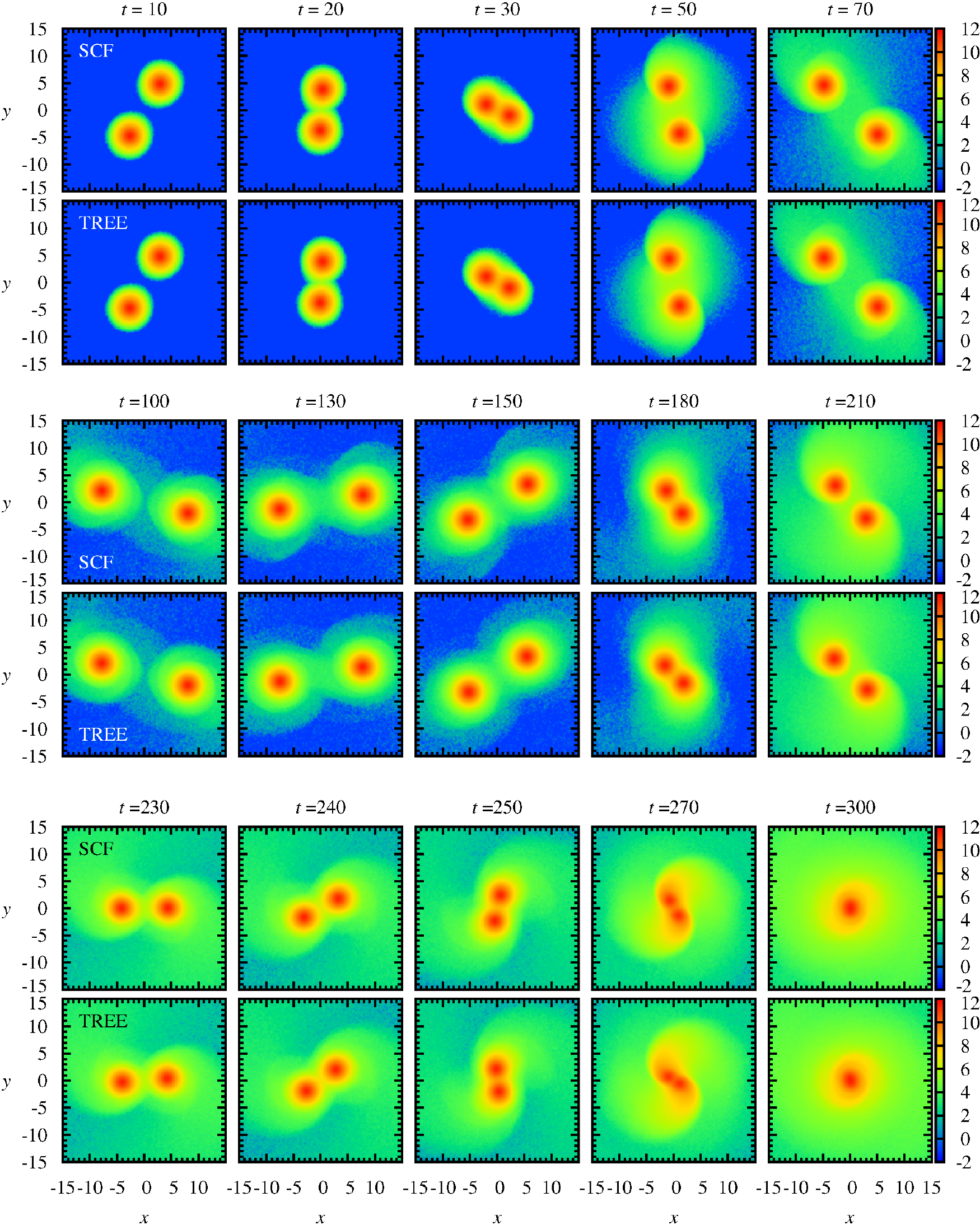}}
\caption{Comparison of the time evolution of the density distribution
of Systems~1 and 2 in the orbital plane for the Case~5 simulations
between the SCF (top panels at each time) and tree code (bottom
panels at each time) simulations.  The density distributions are
represented on a logarithmic scale.}
\label{fig:evolve_12}
\end{figure*}

\begin{figure*}[th]
\centerline{\includegraphics[width=14cm]{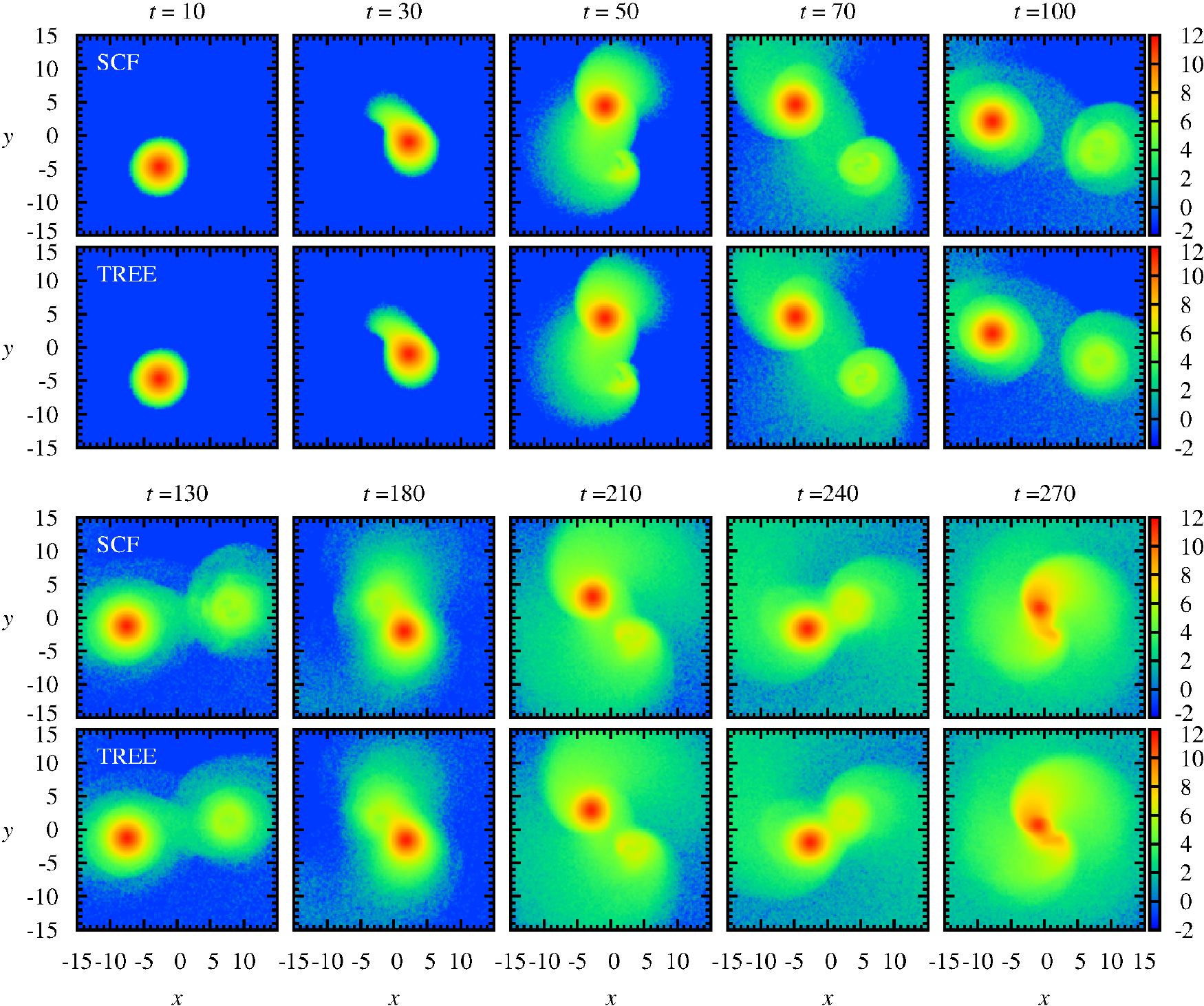}}
\caption{Comparison of the time evolution of the density distribution
of System~1 in the orbital plane for the Case~5 simulations between
the SCF (top panels at each time) and tree (bottom panels at each
time) simulations.  The density distributions are represented
on a logarithmic scale.}
\label{fig:evolve_1}
\end{figure*}

\begin{figure*}[th]
\centerline{\includegraphics[width=15cm]{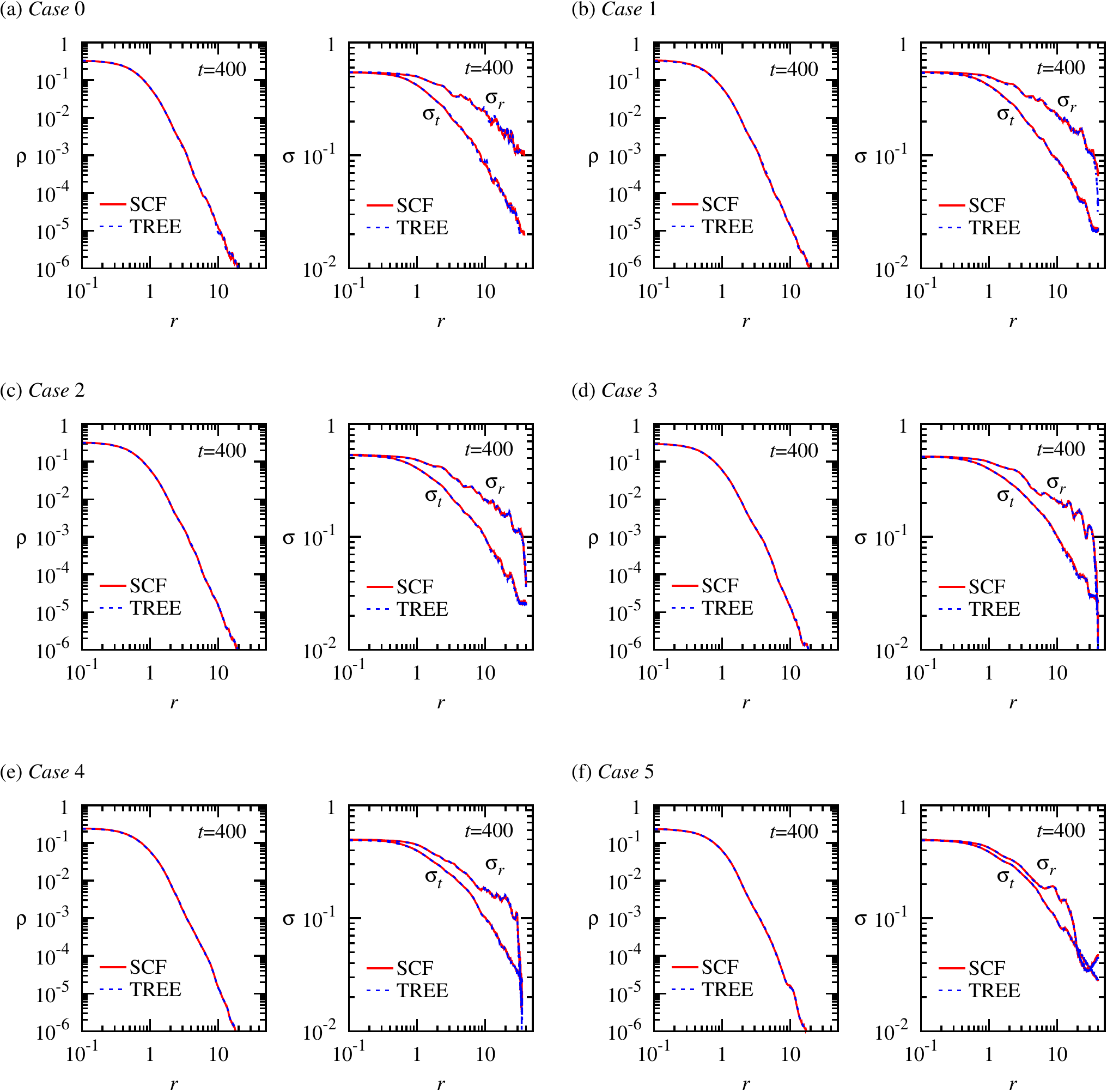}}
\caption{Density and velocity dispersion profiles of the merged
system for (a) Case~0 to (f) Case~5 at $t=400$.  For each case,
the left panel shows the density profile, while the right panel
presents the radial, $\sigma_r$, and tangential, $\sigma_t$,
velocity dispersion profiles.  The red solid and blue dashed
lines correspond to the SCF and tree code simulations, respectively.}
\label{fig:rhovel}
\end{figure*}

\begin{figure*}[th]
\centerline{\includegraphics[width=14cm]{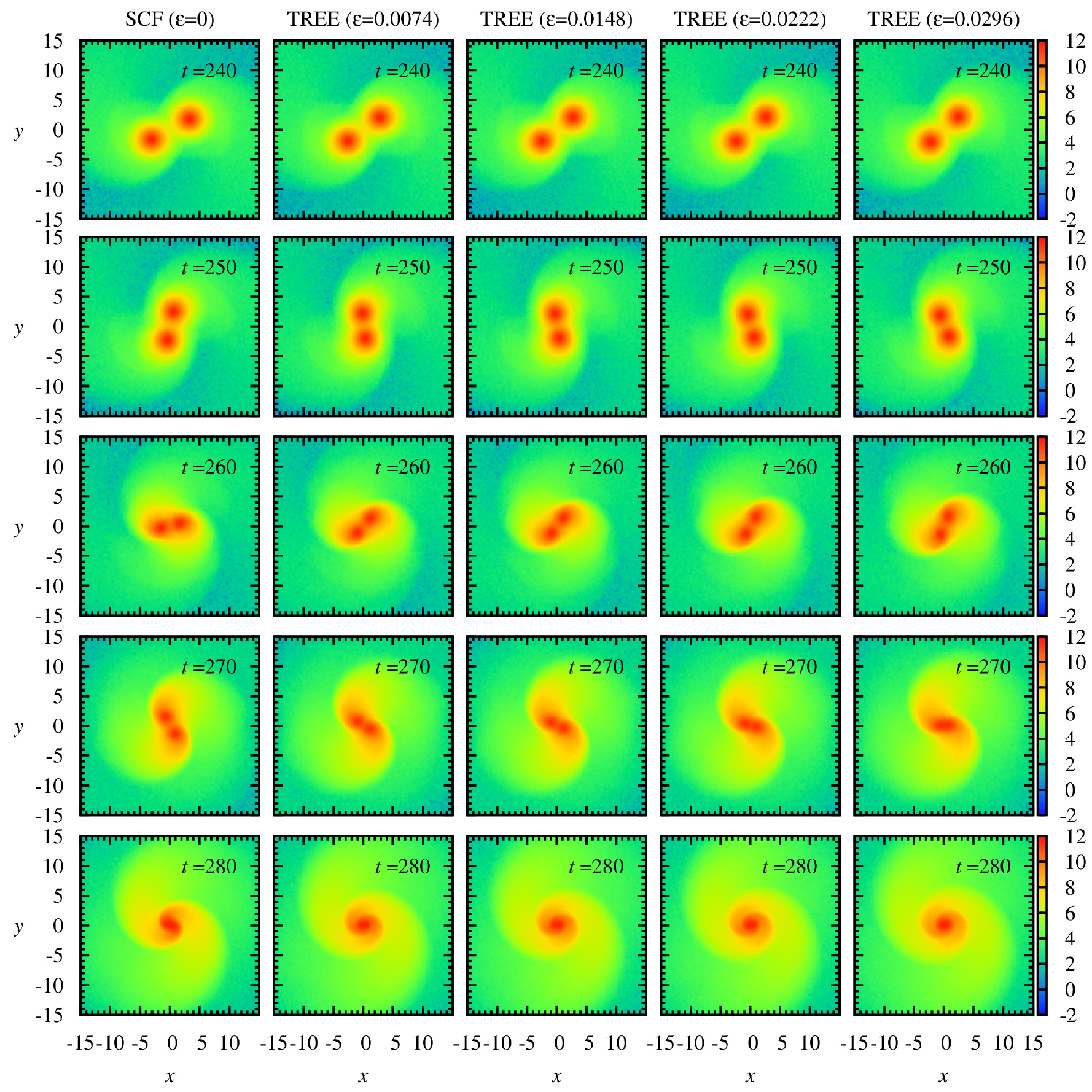}}
\caption{Time evolution of the density distribution of Systems~1
and 2 in the orbital plane for the Case~5 simulations.  The first
column shows the density distributions produced from the SCF simulation
that corresponds to a zero softening length ($\varepsilon=0$).
The second to fifth columns depict those from the tree code simulations
with the Plummer softening lengths of $\varepsilon=0.0074, 0.0148,
0.0222$ and 0.0296, respectively.  As the softening increases,
the orbital phase of the two systems advances faster.  The density
distributions are represented on a logarithmic scale.  The snapshot
time is denoted at the top right corner.}
\label{fig:softening}
\end{figure*}

\begin{figure}[th]
\centerline{\includegraphics[width=8cm]{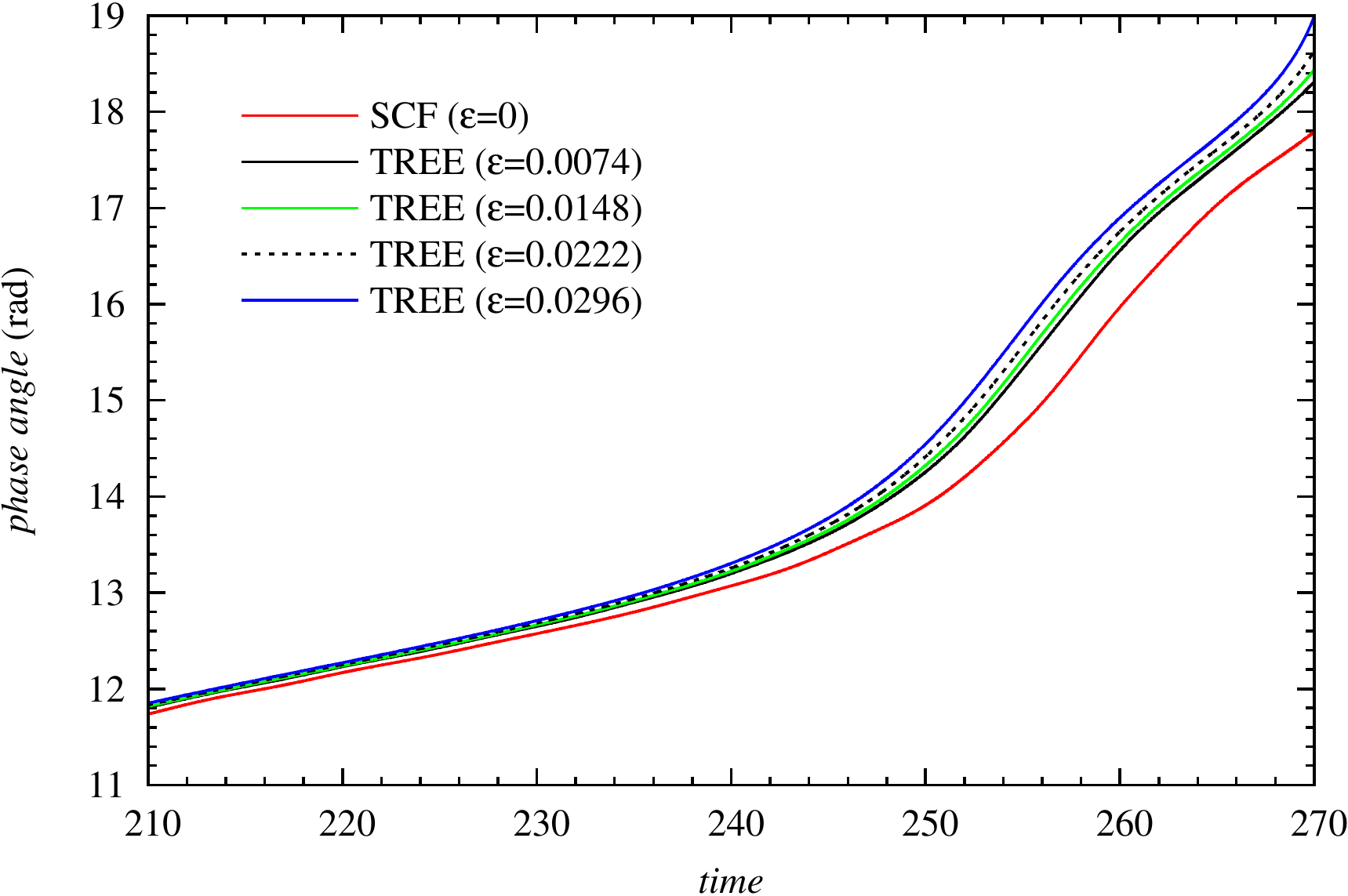}}
\caption{Time evolution of the phase angle for the Case~5
simulations.  The phase angle is defined by the angle between
the $x$-axis and the line connecting to the two centers of
mass of Systems~1 and 2.  The red solid line shows the phase
angle produced from the SCF simulation, while the black solid,
green solid, black dotted, and blue solid lines stand for
that from the tree code simulations with the Plummer softening
lengths of $\varepsilon = 0.0074, 0.0148, 0.0222$, and 0.0296,
respectively.}
\label{fig:phase}
\end{figure}

\begin{figure*}[th]
\centerline{\includegraphics[width=14cm]{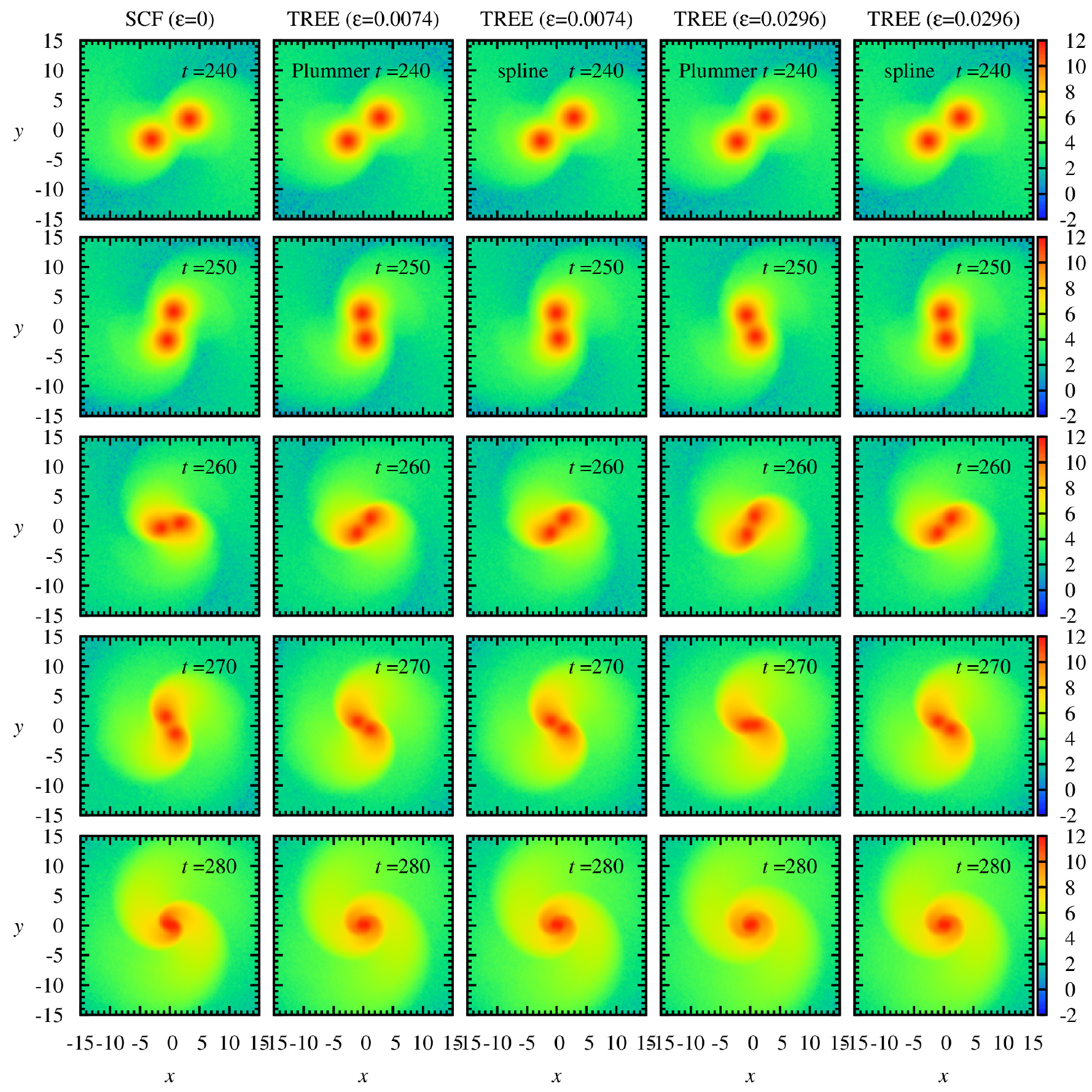}}
\caption{Time evolution of the density distribution of Systems~1
and 2 in the orbital plane for the Case~5 simulations.  The first
column shows the density distributions produced from the softening-free
($\varepsilon=0$) SCF simulation.  The second and third columns
exhibit those from the tree code simulations using the Plummer
and spline softening lengths of $\varepsilon=0.0074$, respectively,
while the fourth and fifth columns depict those using the Plummer
and spline softening lengths of $\varepsilon=0.0296$, respectively.
The density distributions are represented on a logarithmic scale.
The snapshot time is denoted at the top right corner of each panel.}
\label{fig:spline}
\end{figure*}

\begin{figure}[th]
\centerline{\includegraphics[width=8cm]{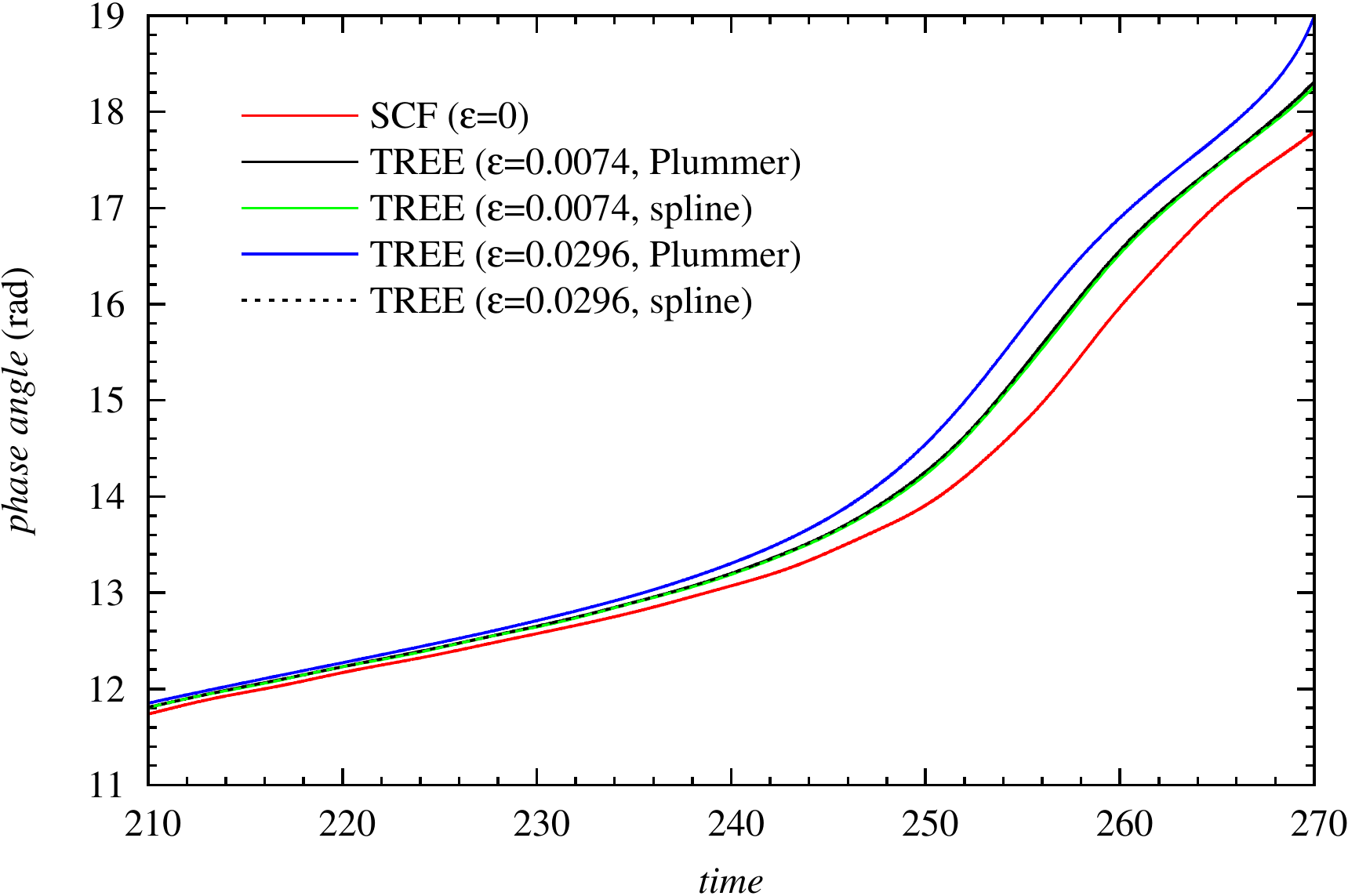}}
\caption{Time evolution of the phase angle for the Case~5
simulations with the tree method using the Plummer and
spline softening lengths, along with the softening-free
SCF method.  The red line shows the phase angle produced
from the SCF simulation.  The black and green solid lines
depict the phase angle with the Plummer and spline softening
lengths of 0.0074, respectively, while the blue solid and
black dotted lines exhibit that with the Plummer and spline
softening lengths of 0.0296, respectively.}
\label{fig:phase_spline}
\end{figure}

\begin{figure}[h]
\centerline{\includegraphics[width=8.5cm]{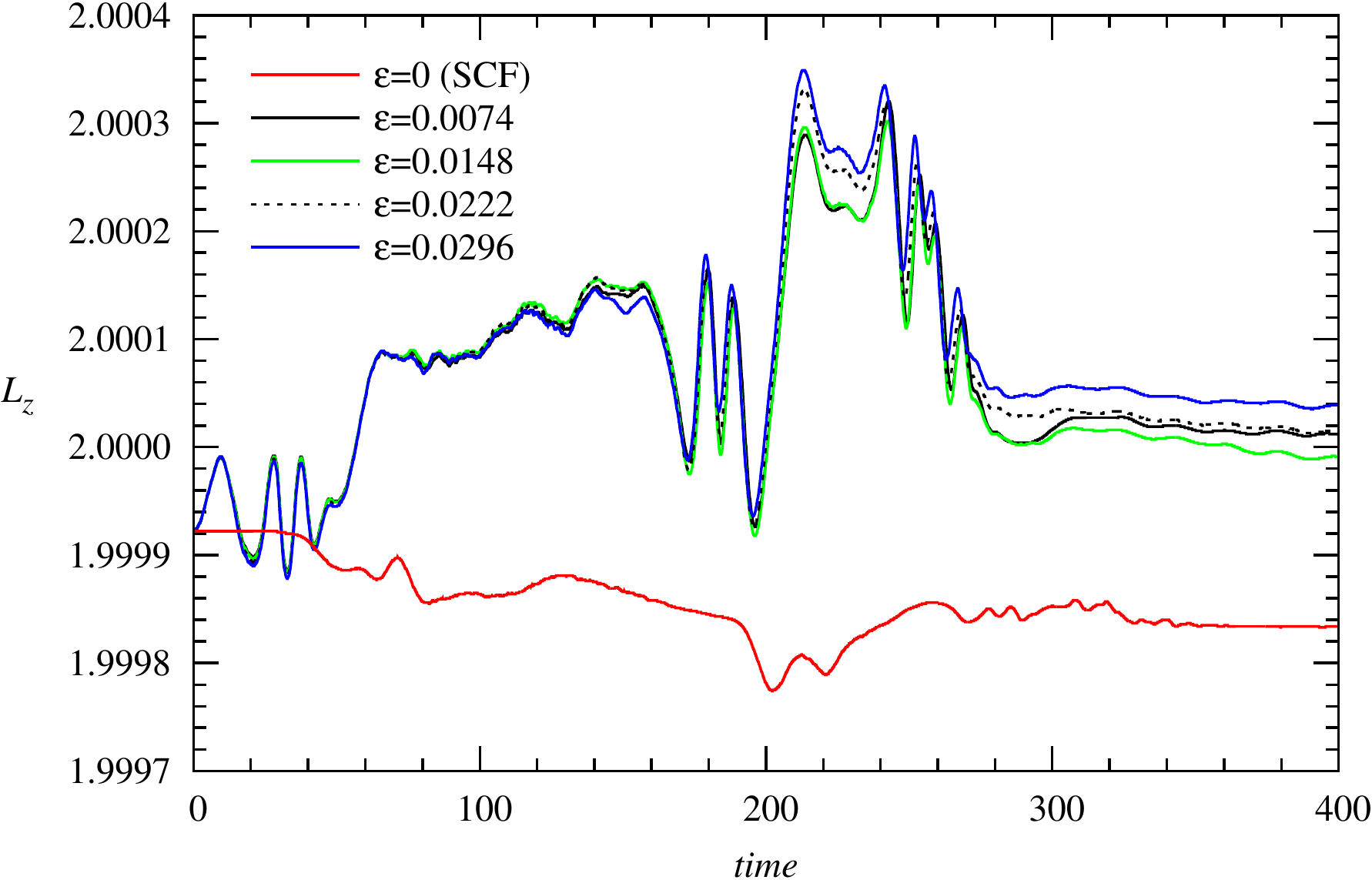}}
\caption{Time change in the $z$-component of the total angular
momentum, $L_z$, for the Case~5 simulations with the tree method
using the Plummer softening, along with the softening-free SCF
method.  The black sold, green solid, black dotted, and blue
solid lines show $L_z$ obtained with the softening lengths of
$\varepsilon=0.0074, 0.0148, 0.0222$, and 0.0296, respectively,
while the red line displays $L_z$ generated by the SCF simulation.}
\label{fig:angmom_plummer}
\end{figure}

\begin{figure}[th]
\centerline{\includegraphics[width=8.5cm]{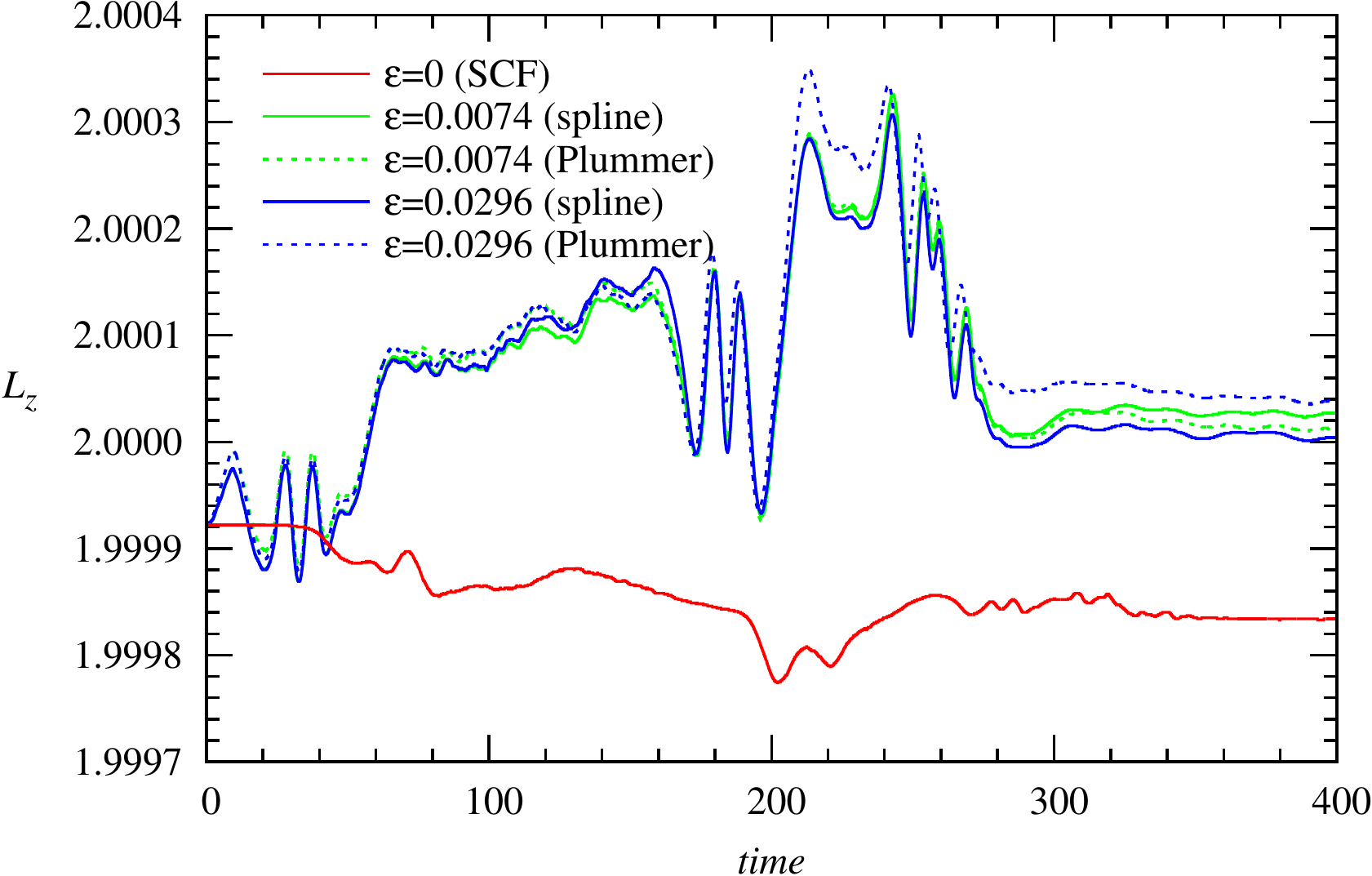}}
\caption{Time change in the $z$-component of the total angular
momentum, $L_z$, for the Case~5 simulations with the tree method
between the spline (solid lines) and Plummer (dotted lines)
softenings, along with the softening-free SCF method.  The
green and blue lines show $L_z$ obtained with the softening
lengths of $\varepsilon=0.0074$ and $\varepsilon=0.0296$,
respectively, while the red line displays $L_z$ generated
by the SCF simulation.}
\label{fig:angmom_spline}
\end{figure}

\section{Results}\label{sec:results}
\subsection{Merging Processes}\label{subsec:processes}
In Figure~\ref{fig:cases0-3}, we compare the orbits of the centers
of mass of Systems~1 and 2 for Cases~0 to 5 between the SCF and
tree code simulations.  The interaction forces between the two
systems in these SCF simulations are calculated using
Equation~(\ref{eq:accel_21}).  We can see that in cases of small
impact parameters for Cases~0 to 3, the orbits of the centers of
mass in the SCF simulations agree quite well with those in the
corresponding tree code ones.  However, in Case~0 which represents
a head-on collision, the fluctuating motions of the centers of
mass in the $y$-direction are much more violent in the SCF simulation
than in the tree code one, although the fluctuating amplitude
is sufficiently small as compared to the size of the merged
system.  In order to demonstrate that this difference originates
from whether the gravitational softening is introduced or not in
tracing the positions of the centers of mass, we calculate them
in the SCF simulation using the same softened gravity of the
Plummer type with $\varepsilon=0.0074$ as is incorporated into
the tree code, instead of using the pure Newtonian force law
described by Equation~(\ref{eq:accel_cm}).  The results are
shown in Figure~\ref{fig:SCF_eps}.  We find from this figure
that the large fluctuating motions of the centers of mass in
the $y$-direction are curbed in the SCF simulation to a significant
degree, and that the fluctuating amplitude is rather smaller
than that in the tree code simulation at late times.  In addition,
the $y$-coordinates of the centers of mass in the tree code
simulation tend to deviate on average almost linearly from
$y=0$ with time, while those in the SCF simulation fluctuate
around $y=0$.  Accordingly, in this case, the linear momentum
of the total system is better conserved for the SCF simulation
than for the tree code one, although in principle, it is not
strictly conserved for both the simulation methods.

Regarding large impact parameters such as Cases~4 and 5,
the orbits of the centers of mass are not in good agreement
between both the simulation methods at late times.  We will
describe the differences in detail below.

From Figure~\ref{fig:cases4-5} (e), we can find that in Case~4,
a noticeable difference exists in the orbits of the centers of
mass of Systems~1 and 2 from $t\sim 200$ to $t\sim 240$ between
the SCF and tree code simulations.  If we adopt the calculation
method of the interaction forces based on
Equation~(\ref{eq:accel_tot_cm}) instead of that based on
Equation~(\ref{eq:accel_21}), desired results are obtained
as presented in Figure~\ref{fig:case4_int} which shows excellent
agreement between the two simulation methods.

Figure~\ref{fig:cases4-5} (f) reveals that in Case~5, the
difference in the orbits of the centers of mass between
the two simulation methods becomes extremely large after
$t\sim 60$.  In addition, the fourth-row panels indicate
that the orbit of the CM in System~2 for the SCF simulation
is rather aligned at late time phases of merging, while
that in System~1 draws a rosette-like feature.  Taking
into consideration the symmetric configuration of the
two identical systems at the beginning, such a spatially
non-uniform distribution of the orbit around the center
of the system should not be expected.  In fact, the tree
code simulation does not exhibit such an alignment tendency
in the orbit.  As shown in Figure~\ref{fig:case5_int}, this
undesired behavior is eliminated by adopting the alternative
calculation method based on Equation~(\ref{eq:accel_tot_cm})
for the interaction forces, and so, the SCF and tree code
simulations agree quite well with each other. 

In Figure~\ref{fig:evolve_12}, we present the time evolution
of the density distributions of Systems~1 and 2 projected
on the orbital plane for Case~5 using the SCF and tree codes.
For the SCF simulation, the interaction forces are calculated
on the basis of Equation~(\ref{eq:accel_tot_cm}).  This figure
demonstrates that the density distribution in the merging two
systems with the SCF method evolves quite similarly to that
with the tree method.  To depict in detail how the individual
systems are tidally deformed in the merging process in this case,
Figure~\ref{fig:evolve_1} illustrates the time evolution of the
projected density distribution of System~1 on the orbital plane
for the SCF and tree code simulations.  We notice that around
$t=50$, a small fraction of particles is stripped off the
main body, subsequently falling into the central region
of System~2.  Figure~\ref{fig:evolve_1} convinces us that
the SCF method can describe the stripping process quite
similarly to that with the tree method, even though the
system is torn apart.

In Figures~\ref{fig:evolve_12} and \ref{fig:evolve_1}, a closer
look at the evolution of the projected density distribution
after $t=240$ for Case~5 reveals that the orbital phase of
the two systems in the tree code simulation advances faster
than that in the SCF simulation.  We will discuss this behavior
in relation to the effects of gravitational softening in the
next section.

\subsection{Merged Systems}
We have found that the two systems in all cases result in almost
completely merged states by $t=400$.  We have also found that
the centers of mass of the two systems continue to oscillate
back and forth around the CM of the total system for long periods,
even though they have merged seemingly into a single system,
but that the oscillation amplitude is sufficiently small as
compared to the size of the merged system as found from
Figure~\ref{fig:cases0-3}.
 
In Figure~\ref{fig:rhovel}, we compare the spherically averaged
density and radial and tangential velocity dispersion profiles
of the merged systems at $t=400$ for Cases~0 to 5 between the
SCF and tree code simulations.  We find from these figures that
both the simulation methods provide almost identical distributions
in the density and radial and tangential velocity dispersions for
the merged system in each case, which implies that the softened
gravity does not affect them at radii sufficiently larger than
the softening length.

\section{Discussion}\label{sec:discussion}
\subsection{Effects of softening length}
We first discuss the finding mentioned in Subsection
\ref{subsec:processes} that the softening length accelerates
the advancement of the orbital phase between the two systems
in the large impact parameter cases.  In order to confirm this
effect, we carry out simulations for Case~5 with the softening
length changed.  Figure~\ref{fig:softening} shows how the
density distribution evolves over time for each value of
$\varepsilon=0.0074,\, 0.0148,\, 0.0222$, and $0.0296$,
while the opening angle of 0.5 is retained, together with
the density distribution produced by the SCF simulation that
corresponds to $\varepsilon=0$.  This figure demonstrates clearly
that the orbital phase advances faster as the softening length
increases.  In Figure~\ref{fig:phase}, we quantify this change
in the orbital phase with time by calculating the angle, $\varphi$,
between the $x$-axis and the line that connects the centers
of mass of Systems~1 and 2 in the orbital plane measured from
$t=0$ at which $\varphi=\pi/4$.  This figure obviously shows
that the larger softening length results in the faster orbital
phase.

We carried out another Case~5 simulation with an opening angle
of 0.3 using $\varepsilon=0.0074$ (not presented here), and
found that there was no practical difference in the orbital
phase between the two opening angles of 0.3 and 0.5 as long
as the same softening length of $\varepsilon=0.0074$ was used.
It therefore follows that the change in the orbital phase is
not caused by the low precision of the force calculation at
large radii.  Moreover, we reran the Case~5 simulations with
$\varepsilon=0.0074$ and 0.0148 by doubling the time step
($\Delta t=0.02$) and confirmed that there were no essential
changes in the orbital phase evolution between the results
with $\Delta t=0.01$ and those with $\Delta t=0.02$ (also not
presented here).  It is thus unlikely that the adopted time step
is the cause of the orbital phase shift.

In order to investigate the origin of the change in the orbital
phase, we carry out Case~5 simulations using spline softening
in which the force law shifts to the pure Newtonian one when
the interparticle distance exceeds twice as large as the
softening length \citep{hk89}.  Results are summarized in
Figure~\ref{fig:spline} where the comparisons between the
Plummer and spline softenings are made for $\varepsilon=0.0074$
and 0.0296.  In either case of $\varepsilon$, the spline softening
delays the orbital phase as compared to the corresponding Plummer
softening.  In fact, Figure~\ref{fig:phase_spline} shows that the
orbital phase for the \mbox{Plummer} softening advances faster than that
for the spline softening if the same softening length is used.
We thus infer that the advancement of the orbital phase depends
on the extent of the deviation from the pure Newtonian force.
Although the spline softening confines the extent of the softened
force to a certain distance, it still accelerates the orbital phase
faster than that in the SCF simulation.

In Figure~\ref{fig:angmom_plummer}, we show how the conserved
quantity of the $z$-component of the total angular momentum, $L_z$,
changes with time for the Case~5 tree code simulations using
the Plummer softening, together with $L_z$ for the corresponding
SCF simulation.  We can see that the value of $L_z$ begins to
increase rapidly at $t\sim 210$ and remains large until $t\sim 260$,
regardless of the softening lengths.  Furthermore, the value of
$L_z$ becomes larger as the softening length increases for the
period between $t\sim 210$ and $t\sim 260$.  Since the orbital
angular momentum could contribute to the total angular momentum
to a considerable degree in this case, the increase in $L_z$
would lead to a faster orbital motion, particularly at the
times after $t\sim 240$ when the distance between the centers
of mass of the two systems becomes smaller and smaller.  From
these circumstances, we could infer that the orbital phase
advances faster as the softening length becomes larger for
that period.  Figure~\ref{fig:angmom_spline} shows the time
change in $L_z$ for the Plummer and spline softening lengths
of $\varepsilon=0.0074$ and 0.0296.  This figure indicates
that the time change in $L_z$ for the spline softening lengths
of $\varepsilon=0.0074$ and 0.0296 is similar to that with
the Plummer softening length of $\varepsilon=0.0074$.
Consequently, we could understand that the time evolution
of the orbital phase angle for the spline softening lengths
of $\varepsilon=0.0074$ and 0.0296 behaves similarly to that
with the Plummer softening length of $\varepsilon=0.0074$, as
shown in Figure~\ref{fig:phase}.  However, since the increase
in $L_z$ is at most on a level of $0.02$\%, it is necessary to
examine further whether such a slight increase can justify this
reasoning.  On the other hand, $L_z$ is well-conserved in the
SCF simulation, as is presented in Figures~\ref{fig:angmom_plummer}
and \ref{fig:angmom_spline}.  This fact suggests that the SCF
simulation could describe the true evolution of the merging
process in this case more precisely than the tree code simulations.
In fact, concerning the orbital phase, the SCF simulation appears
to correspond to a limiting case of $\varepsilon=0$ when the
softening length is decreased from $\varepsilon=0.0296$ to
$\varepsilon=0.0074$ in the tree code simulations.

In order to reveal how faithfully the SCF simulations shown
here can follow the true evolution, the most desirable approach
may be a comparison of the simulations with the SCF code to
those with a six-dimensional phase-space code developed by
\citet{yyu13}, which has been touched up by \citet{tymy17},
or with that based on a moving adaptive simplicial tessellation
method devised by \citet{sc16}, since no gravitational softening
is included in these codes.  In fact, \citet{yyu13} have
presented a sample simulation of merging two identical King
spheres.  This implies that the comparison mentioned above
is feasible.

\subsection{Computation time}
The SCF code used here has the perfect scalability \citep{hsb95},
so that ideal
load-balancing is easily realized on a massively parallel computer.
As a consequence, on such a machine, the cpu time of an SCF simulation
is proportional to
$N_{\rm core}(n_{\rm max}+1)(l_{\rm max}+1)(m_{\rm max}+1)$, 
where $N_{\rm core}$
is the number of particles per cpu core.  On the other hand,
the perfect scalability cannot be realized for tree algorithms.
Furthermore, in the tree code used for the current study on which
the FDPS library developed by \citet{iwasawa16} has been implemented,
an exchange of particles between computing nodes is executed
for efficient load-balancing as an additional computation cost.
Consequently, when 648 cpu cores are operated on a Cray XC30 system,
the cpu times for the SCF simulations are $\sim 3.2\times 10^{-4}$
sec per step per core when Equation~(\ref{eq:accel_21}) is
adopted, and they are $\sim 2.5\times 10^{-3}$ sec per step
per core when Equation~(\ref{eq:accel_tot_cm}) is employed,
while those for the tree code simulations are $\sim 4.8\times
10^{-3}$ sec per step per core.  Thus, we can see that the SCF
code is more than an order of magnitude faster than the tree
code, when the interaction forces between the two systems are
based on Equation~(\ref{eq:accel_21}), and that the former is
at least twice as fast as the latter, even though the interaction
forces are calculated according to Equation~(\ref{eq:accel_tot_cm}).

As \citet{mlhs14} have implemented an SCF code for simulating
isolated stellar systems on a machine carrying a GPGPU which
deals with parallel computation, the present SCF code can also
be adjusted to such a machine without any difficulty.  It will
therefore make feasible merging simulations with a sufficiently
large number of particles on a GPGPU system within a relatively
small amount of computation time.

\subsection{Applications of the SCF code for merging}
Since an SCF method is literally a field method, we can obtain
a force field at each time step as an SCF simulation proceeds
with time.  In addition, the force field is represented by the
expansion coefficients, $A_{nlm}$, so that even though $A_{nlm}$
are saved at each time step, the data size is extremely small
as compared to that when snapshots are saved at each time step.
This means that the orbit of an arbitrary particle in phase
space can be traced once an SCF simulation has been completed,
because the acceleration at an arbitrary position can be
calculated at each time step from $A_{nlm}(t)$ as indicated
by Equation~(\ref{eq:acc_basis}).  Therefore, phase space at
a specified time can be reproduced by tracing the orbits of
necessary particles at time $t$ backward to $t=0$ on the basis
of Liouville's theorem, as has been demonstrated by \citet{sh97}.
In particular, if we analyze the structures in phase space
reproduced from a merging simulation, traces of the merging
process that the Milky Way experienced in the past might be
found out by comparing them to those derived from the Gaia
data, just as \citet{antoja18} have recently attacked this
line of study.

The advantage of the present SCF code that the force field
of two merging spherical galaxies is made available at each
time step can be leveraged to search for the parameters that
reproduce the observed tidal features of interacting disk
galaxies.  This line of investigation has been realized as
Identikit 1 by \citet{bh09}, and as Identikit 2 by \citet{josh11}.
In these programs, a simulation is carried out for two interacting
self-gravitating dark halos each of which contains a galactic
disk represented by test particles.  If a tree algorithm is
incorporated into Identikit 1 or 2, we have to carry out a
simulation each time we change the configurations of the disks
that are immersed in the halos, even though the configuration
of the two halos remains unchanged.  This is because the
force field of two interacting halos cannot be found easily
as long as a tree code is used.  On the other hand, once an
SCF simulation for two interacting halos has been carried
out, even though the disk configurations are changed, we
do not need to repeat such a simulation, because we have
already known the time-dependent force field that the two
halos created.

Another advantage that the SCF code can execute a sizable
simulation in a relatively short computation time will make
feasible the merging of two spherical galaxies each of which
has a supermassive black hole (SMBH) at the center.  For such
simulations, an extremely small time step is required to cope
with a wide dynamic range originating from the existence of the
SMBH.  In addition, it is desirable to assign as many particles
as possible to each galaxy to investigate the detailed structures
down to small radii around the SMBH.  These circumstances
unavoidably force us to require a long computation time.
\citet{me96} and \citet{makino97} conducted some pieces of
pioneering work concerning the simulations mentioned above
using a special-purpose computer called a GRAPE-4 system
\citep{taiji96} on which the NBODY1 code \citep{sa85} was
optimized to run, and disclosed that the resulting cusps
were similar to those found observationally \citep{lauer95,
richstone96}.  Unfortunately, because of a challenging issue
at that time, they managed to assign at most hundreds of
thousands of particles to a galaxy.  The number of particles
available to the SCF code developed here would be roughly two
or three orders of magnitude larger than that they used, so
that we can expect to unravel the detailed properties of the
cusp formed from the above-noted merging.

\section{Conclusions}\label{sec:conclusions}
We have developed an SCF code for merging two spherical stellar
systems.  This SCF code has been evaluated by comparing the results
of merging simulations it generates to those obtained using a tree
code for a wide range of impact parameters.  We have confirmed
that the results generated by the SCF code are in excellent
agreement with those by a tree code: the similarity is found
in the evolution of the density distribution in the merging
processes, and in the density and radial and tangential velocity
dispersion profiles of the merged systems.

The softening-free SCF code has revealed that the softening length
of the Plummer type used in the tree code causes the advancement of
the orbital phase of the two interacting systems for large impact
parameters.  In addition, the orbital phase becomes faster as the
softening length increases.  We have demonstrated that the faster
advancement is caused by the larger convergence length to the pure
Newtonian force.

Since the SCF code generates the force field at each time step
through the expansion coefficients, we can reproduce phase space
on the basis of Liouville's theorem.  Furthermore, if the SCF
code is incorporated into Identikit 1 or 2, the force field
is known from a merging simulation, so that we can give an
efficient means for finding the parameters that lead to the
observed morphology of interacting disk galaxies.

The present SCF code is at least twice as fast as a tree code,
so that a suitable application could be the merging of two
spherical galaxies each of which contains an SMBH at the center
using a sufficiently large number of particles.

\acknowledgments
We acknowledge Prof.~J.~E.~Barnes for suggesting an application
of the SCF code for merging simulations, and Prof.~L.~Hernquist
for his careful reading of the manuscript.  We are indebted to
Prof.~J.~Makino for his valuable comments on the effects of
gravitational softening.  Thanks are also due to Prof.~K.~Takahashi
for providing us with his program for constructing King models.
The SCF and tree code simulations were carried out on the Cray
XC30 and XC50 systems at the Center for Computational Astrophysics
at the National Astronomical Observatory of Japan.

\appendix

\section{The density and potential basis functions}\label{app:A}
We employ the basis set constructed originally by \citet{cb73}
and reformulated by \citet{ho92}.  %\citep[see also][]{ho92}.
The density and potential basis functions, respectively,
denoted by $\rho_{nlm}(\textbf{\textit{r}})$ and
$\Phi_{nlm}(\textbf{\textit{r}})$, are  represented
in polar coordinates by
\begin{equation}
\rho_{nlm}(\textbf{\textit{r}})=\frac{K_{nl}}{4\pi}
\frac{a^{l+5}\,r^l}{{(a^2+r^2)}^{l+5/2}}C_{n}^{(l+1)}(\xi)
\sqrt{4\pi}\,Y_{lm}(\theta,\,\phi),
\label{eq:density_basis}
\end{equation}
and
\begin{equation}
\Phi_{nlm}(\textbf{\textit{r}})=-\frac{a^{l+1}\,r^l}{{(a^2+r^2)}^{l+1/2}}
C_{n}^{(l+1)}(\xi)\sqrt{4\pi}\,Y_{lm}(\theta,\,\phi),
\label{eq:potential_basis}
\end{equation}
where $a$ is the scale length, $C_n^{(\alpha)}(\xi)$ are the
ultraspherical, or Gegenbauer polynomials \citep{as72} with $\xi$
being the radial transformation written by
\begin{equation}
\xi=\frac{r^2-a^2}{r^2+a^2},
\end{equation}
and $Y_{lm}(\theta,\,\phi)$ are the spherical harmonics that are
related to associated Legendre polynomials, $P_{lm}(\cos\theta)$, by
\begin{equation}
Y_{lm}(\theta,\,\phi)=\sqrt{\frac{2l+1}{4\pi}\frac{(l-m)!}{(l+m)!}}
\,P_{lm}(\cos\theta)\exp(i\,m\phi).
\end{equation}

In Equation (\ref{eq:density_basis}), the normalization factor,
$K_{nl}$, is expressed by
\begin{equation}
K_{nl}=4n(n+2l+2) + (2l+1)(2l+3).
\end{equation}

\end{document}